\documentclass[10pt,english,notitlepage,a4paper,superscriptaddress,floatfix,longbibliography,prl,twocolumn]{revtex4-2}

\usepackage[T1]{fontenc}
\usepackage{babel}
\usepackage{amsmath}
\usepackage{amssymb}
\usepackage{graphicx}

\usepackage{bm}
\usepackage{ulem}
\usepackage[unicode=true,pdfusetitle,
urlcolor=blue,bookmarks=true,bookmarksnumbered=false,bookmarksopen=false,
 breaklinks=false,pdfborder={0 0 1},backref=false,colorlinks=true]
 {hyperref}
\usepackage{comment}
\usepackage[percent]{overpic}

\makeatletter

\newcommand{\region}[1]{(\textbf{\lowercase{#1}})}

\usepackage{xcolor}
\usepackage{tikz}
\definecolor{myblue}{RGB}{12, 12, 158}
\definecolor{myred}{RGB}{158, 19, 22}
\definecolor{myorange}{RGB}{245, 150, 12}
\definecolor{mygreen}{RGB}{26, 148, 49}
\definecolor{Prune}{RGB}{99,0,60}
\definecolor{Purple}{RGB}{75, 0, 130}
\definecolor{Pink}{RGB}{255, 105, 180}
\definecolor{deepskyblue}{RGB}{0, 191,255}
\definecolor{limegreen}{RGB}{50, 205, 50}
\definecolor{crimson}{rgb}{0.86, 0.08, 0.24}

\definecolor{blue(ncs)}{rgb}{0.0, 0.53, 0.74}

\makeatother

\begin{document}
\title{The Copycat Perceptron: Smashing Barriers Through Collective Learning}

\author{Giovanni Catania}
\email{gcatania@ucm.es}
\affiliation{Departamento de Física Teórica, Universidad Complutense de Madrid,
28040 Madrid, Spain}
\author{Aur\'{e}lien Decelle}
\affiliation{Departamento de Física Teórica, Universidad Complutense de Madrid,
28040 Madrid, Spain}
\affiliation{Université Paris-Saclay, CNRS, INRIA Tau team, LISN, 91190 Gif-sur-Yvette,
France.}
\author{Beatriz Seoane}
\affiliation{Université Paris-Saclay, CNRS, INRIA Tau team, LISN, 91190 Gif-sur-Yvette,
France.}
\begin{abstract}
We characterize the equilibrium properties of a model of $y$ coupled binary perceptrons in the teacher-student scenario, subject to a suitable cost function, with an explicit ferromagnetic coupling proportional to the Hamming distance between the students' weights. In contrast to recent works, we analyze a more general setting in which thermal noise is present that affects each student's generalization performance. In the nonzero temperature regime, we find that the coupling of replicas leads to a bend of the phase diagram towards smaller values of $\alpha$: This suggests that the free entropy landscape gets smoother around the solution with perfect generalization (i.e., the teacher) at a fixed fraction of examples, allowing standard thermal updating algorithms such as Simulated Annealing to easily reach the teacher solution and avoid getting trapped in metastable states as it happens in the unreplicated case, even in the computationally \textit{easy} regime of the inference phase diagram. These results provide additional analytic and numerical evidence for the recently conjectured Bayes-optimal property of Replicated Simulated Annealing (RSA) for a sufficient number of replicas.
From a learning perspective, these results also suggest that multiple students working together (in this case reviewing the same data) are able to learn the same rule both significantly faster and with fewer examples, a property that could be exploited in the context of cooperative and federated learning.
\end{abstract}

\maketitle

Statistical mechanics provides valuable tools for understanding machine learning. The goal is to describe the typical behavior of a neural network with respect to global parameters such as the training set size~\citep{gardner1988optimal,sompolinskylearningfromexamples} or the gradient descent noise~\citep{zdeborova2016statistical,mannelli2020marvels}. This approach helps us to understand the conditions that favour a better learning and when achieving good performance is impossible.
An impressive example of this approach's effectiveness is the analysis of the solution space as a function of the proportion of clauses in $k$-SAT in classical combinatorial optimization problems~\citep{mezard_analytic_2002,zdeborova2007phase,krzakala2009hiding}.\\At zero temperature, accurately classifying labeled data with random labels can be seen as a constraint satisfaction problem (CSP). The examples range from perceptrons~\citep{gardner1988optimal,krauth1988basins} to more complex architectures such as the Committee Machine~\citep{monasson1995weight,monasson1995learning}, Support Vector Machines~\citep{oppersvm}, multilayered perceptrons~\citep{barkaiandSompmultilayered,Cornacchia_2023}, or even  continuous optimization problems in condensed matter~\cite{Franz_2016}. Surprisingly, despite the vast number of potential solutions with poor generalization, perceptrons and deep neural networks excel in classification tasks. This suggests that the standard training methods do not explore the entire space of quasi-optimal CSP solutions, but use a more efficient approach.\\ In recent years, researchers have introduced a theoretical framework based on the concept of \textit{local entropy} to better understand the effectiveness of training algorithms and to find solutions that generalize well. This concept, along with the coupled-replicas strategy, has been thoroughly investigated in several studies~\citep{baldassilocal2016, baldassisubdominant2015, baldassi_unreasonable_2016, huang_entropy_2013}, also in its connection to quantum annealing protocols \citep{baldassiQuantumAnnealing}. Moreover, a recent work~\citep{angelini_limits_2023} has provided convincing evidence that coupling replicas can help in identifying favourable local minima and avoiding entrapment in glassy states, as shown in the graph coloring problem.\\ In this study, we focus on the paradigmatic binary perceptron model in the \textit{teacher-student} scenario, a well-establised example of a planted inference problem ~\citep{sompolinskylearningfromexamples,barthel2002hiding,zdeborova2016statistical} where a \textit{student} perceptron attempts to learn the classification rule from the examples given by a \textit{teacher} perceptron. In this setting, the ratio between the number of examples and the number of parameters ($\alpha$) acts as a signal-to-noise ratio. \\Previous studies have shown that this model, unlike its continuous counterpart, exhibits a first-order phase transition at zero temperature, corresponding to a sharp decrease in the generalization error as the number of training examples increases~\citep{gyorgi_binaryperceptron, sompolinskylearningfromexamples, statmechlearningruleBiel}.
Recent works have focused on describing this storage performance at zero temperature in the so-called \textit{robust ensemble}~\citep{baldassisubdominant2015, huang_entropy_2013, huang_origin_2014}, where multiple replicas of the same model interact through a ferromagnetic coupling to favour solutions with high local entropy. This ensemble is closely related to the Franz-Parisi potential in glassy physics~\citep{monasson1995structural,franz:jpa-00247146}. Similarly, the idea of coupled-replica models has been used to design efficient cooperative local search strategies in simple neural networks \citep{Huang_2011}, or in deep learning contexts ~\citep{NIPS2015_d18f655c}. \\ In this paper, we propose a more general theoretical approach in which we derive the entire $(\alpha,T)$ phase diagram of the binary perceptron in the robust ensemble as a function of the number of coupled replicas $y$ or the coupling $\gamma$. This phase diagram can be used to understand how the structure of the solution space evolves during the training process and how the $\gamma$ coupling favors broader and flatter landscapes.
We show that the phase diagram allows us to propose more effective annealing strategies, especially in non-Bayesian optimal scenarios. Moreover, we show that the so-called inference-\textit{easy} region is plagued by subdominant 1RSB metastable states that make it practically impossible for annealing strategies to find the teacher in a reasonable time. Using the phase diagram obtained with the replica trick, we show that coupled perceptrons can easily learn the teacher solution in this region by avoiding the 1RSB states.

 \textit{Model -- } The model we consider is defined by a number $y$ of binary perceptron students and a teacher perceptron. Each student $u\in\left\{ 1,\ldots,y\right\} $ is parameterized by a weight vector with $N$
components $\boldsymbol{w}_{u}=\left\{ w_{i,u}\right\} _{i=1}^{N}$
with $w_{i,u}\in\left\{ -1,1\right\} $.
Students are given the same set of $M$ labeled data points $\mu$, each represented by a pair $\left(\boldsymbol{\xi}^{\mu},\sigma_{0}^{\mu}\right)$:
the first is the {\it training sample} $\boldsymbol{\xi}^{\mu}$, an $N$-dimensional vector with binary entries $\xi_{i}^{\mu}\in\left\{ -1,1\right\} $,
and $\sigma_0^{\mu}$ is the corresponding {\it label}, assigned by a \textit{teacher} perceptron with weight vector $\bm{w}_0$. The input-output relation for both the teacher and the students is defined as follows:
\begin{align}
\sigma_{u}^{\mu} & =\text{sign}\left(\omega_{u}^{\mu}\right) \text{ and } \omega_{u}^{\mu} =\frac{\boldsymbol{w}_{u}\cdot\boldsymbol{\xi}^{\mu}}{\sqrt{N}}\label{eq:def_output_teacher}, u\in \{0,\dots,y\} 
\end{align}
for $\mu\in\left\{ 1,\ldots,M\right\} $, where the symbol $\cdot$ denotes the scalar product and $u=0$ refers to the teacher. The normalization factor in Eq. \eqref{eq:def_output_teacher} ensures that the energy of the model is extensive in the system size $N$. At fixed planted configuration, the labels generated by the teacher are noiseless according to \eqref{eq:def_output_teacher} so that the Bayes-optimal temperature is $T=0$.

Each student adjusts their weights to minimize an appropriate cost function, which accounts for
 the number of misclassified examples, i.e. those where the assigned label differs from that of the teacher. Moreover, the students will interact through a pairwise ferromagnetic coupling that favours configurations in which students have a high mutual overlap. In other words: Imitation between students is encouraged.

The learning is defined by specifying the Hamiltonian of the system. Following~\citep{engel_statistical_2001}, we first define the \textit{stability parameters} as
$\Delta_{u}^{\mu} = \sigma_{0}^{\mu} \omega_u^{\mu}$,
so that $\Delta_{u}^{\mu} > 0$ whenever the input $\mu$ is correctly classified by the student $u$.
Secondly, we define an arbitrary potential $V\left(\Delta\right)=\left(-\Delta\right)^\nu \Theta\left(-\Delta\right)$,
where $\nu \in \mathbb{Z}^+$ and $\Theta\left(x\right)$ denotes the Heaviside step function.
The potential assigns $0$ energy to the weight vectors that correctly classify a given example (i.e. those for which  $\Delta_u^{\mu} > 0$) and a positive cost $\propto \left(-\Delta\right)^{\nu}$ to  misclassified ones (i.e when $\sigma_u^{\mu} \neq \sigma_0^{\mu}$).\\
The Hamiltonian of the model can be written as:
\begin{equation}
\mathcal{H}\left(\left\{ \boldsymbol{w}_{u}\right\} _{u=1}^{y} \! \mid \!\boldsymbol{w}^{0},\mathbb{D}\right)\!=\!
\sum_{u,\mu}^{y,M} V\left(\Delta_{u}^{\mu}\right)-\frac{\gamma}{\beta  y}\sum_{u<v}\boldsymbol{w}_{u}\cdot\boldsymbol{w}_{v},\label{eq:HamiltonianCoupledPerceptrons}
\end{equation}
where $\beta=T^{-1}$ and $\mathbb{D}=\left\{ \left(\boldsymbol{\xi}^{\mu},\sigma_{0}^{\mu}\right) \right\} _{\mu=1}^{M}$ denotes the training set of labeled examples. The second term quantifies the interaction between student pairs and it is proportional to the Hamming distance between their weights. Eq.~\eqref{eq:HamiltonianCoupledPerceptrons}
describes a system in which $y$ students try to learn the same teacher's rule, while interacting with a ferromagnetic potential (tuned by $\gamma$): The latter favors configurations in which students have a similar $\bm w$, i.e., students are actively encouraged to be ``inspired'' by their peer perceptrons.
Different choices of the exponent $\nu$ lead to distinct learning dynamics, each with unique convergence properties and non-zero temperature phase diagrams in the thermodynamic limit~\citep{horner_dynamics_1992,Horner1992_DMFT,sompolinskylearningfromexamples}. In this article, we will focus on the case $\nu\!=\!1$, the so-called ``perceptron rule", originally introduced by Rosenblatt in~\citep{rosenblattPerceptron} for continuous weights. The case $\nu\!=\!0$ is less interesting as the system remains frozen at all temperatures~\cite{horner_dynamics_1992}; the case $\nu\!\geq\! 2$ will be the subject of future studies.

We consider a scenario where each student fluctuates with a heat bath at a given $\beta$ and a fixed interaction strength between students, $\gamma$. We seek to characterize whether and how the different operating regimes of the binary perceptron change when considering multiple interacting replicas. In other words, our goal is to understand and model how the collective or cooperative learning of multiple students differs at a fundamental level from that of a single student.

\noindent \textit{Mean-field theory --} To characterize the equilibrium properties of the model~\eqref{eq:HamiltonianCoupledPerceptrons}, we need to compute the quenched free entropy in the thermodynamic limit where both $N, M\!\to\! \infty$ keeping the ratio $\alpha\!=\!M \slash N$ finite.
The set of patterns $\left\{ \boldsymbol{\xi}^{\mu}\right\} _{\mu=1}^{M}$
and the teacher's weight $\boldsymbol{w}_0$ represent the quenched disorder to average over. As each student reviews the same examples, the disorder is the same for all students.
The starting point of the derivation is the partition function $Z\!=\!\sum_{\boldsymbol{w}_{u}} e^{-\beta \mathcal{H}}$. In our definition for the Hamiltonian ~\eqref{eq:HamiltonianCoupledPerceptrons}, the thermal noise only affects the first term, but not the coupling between the students: this choice, consistent with previous works \citep{angelini_limits_2023}, allows a finite $\gamma$ between the replicas even in the $T\to 0$ limit, where the model is reduced to a CSP in the space of students' weights that satisfies the constraints imposed by the teacher. 
The physical quantity of interest is the quenched free entropy density \begin{equation}
    \mathcal{G}(\alpha, \beta, \gamma,y) = \lim_{N\to \infty} \frac{1}{yN}\left\langle \log Z \right\rangle_{\mathbb{D}, \boldsymbol{w}^0}.\label{eq:freeenergy}
\end{equation}
For simplicity, we restrict ourselves to finite values of $y$, although in principle the limit $y \to \infty$ could also be analyzed, since the coupling between the students is rescaled so that \eqref{eq:freeenergy} is an intensive quantity with respect to $y$. The computation of \eqref{eq:freeenergy} can be performed using the usual  {\it replica trick} in spin-glass theory~\citep{mvpbook,charbonneau2023spin}. The main difficulty here lies in dealing with the coupled students. Formally, they are ``real'' replicas: they share the same quenched disorder as the replicas of the replica trick, but they also interact through an explicit pairwise coupling.
The Mean-field theory requires the introduction of a set of order parameters, to be evaluated with the saddle-point method in the thermodynamic limit. These are the overlap each student has with the teacher vector (in each replica $a\in\left\{1,\ldots,n \right\}$) and the pairwise overlap between two students $u,v$ in replicas $a,b$ respectively:
\begin{align}
    R_{a}^{u} & =\frac{1}{N}\sum_{i}w_{i,u}^{\left(a\right)}w_{i}^{0} \quad\text{and}\quad q_{ab}^{uv} =\frac{1}{N}\sum_{i}w_{i,u}^{\left(a\right)}w_{i,v}^{\left(b\right)}.\label{eq:R_au_def-1}
\end{align}
The simplest Replica Symmetric (RS) ansatz imposes a permutation symmetry between the overlaps in the replica space, i.e. $q_{ab}^{uv}\!=q$, $\forall a\!\neq\! b$.
The diagonal blocks $a\!=\!b$ represent the average correlation between the students, which in general may depend on the topology of the interactions in the replicated space. Assuming a fully-connected topology as in \eqref{eq:HamiltonianCoupledPerceptrons}, it is natural to assume a uniform overlap between students $u\!\neq\! v$ in the \textit{same} replica $a$, so that $q_{aa}^{uv} \!=\! \delta_{uv} + p \left(1 \!-\! \delta_{uv}\right)\; \forall a$
(the diagonal constraint $\delta_{uv}$ comes from the binary nature of the weights).
Due to the different nature of the replicas $a,b$ and the students $u,v$, the two values of the overlap used in this ansatz are expected to be different: in particular, $p\geq q$ due to the ferromagnetic coupling $\gamma$ in the Hamiltonian (the equality holds at $\gamma\!=\!0$). As far as the \textit{signal} term $R_a^u$ is concerned, the RS ansatz implies $R_a^u\!=\!R \; \; \forall u,a$. A complete derivation of the explicit form of $\mathcal{G}$ in~\eqref{eq:freeenergy} under this ansatz is given in Appendix~\ref{app:MFTheory}. In the thermodynamic limit, the equilibrium behavior of the model is determined by the maxima of this free entropy, which are found by imposing stationarity w.r.t. the order parameters. In particular, the teacher configuration (i.e. the planted solution) corresponds to a free-entropy maximum with $R\!\simeq\! 1$ (the strict equality holding at $T\!\to\! 0$).
\begin{figure}[t!]
\begin{overpic}[trim=0 10 0 5,width=\columnwidth]{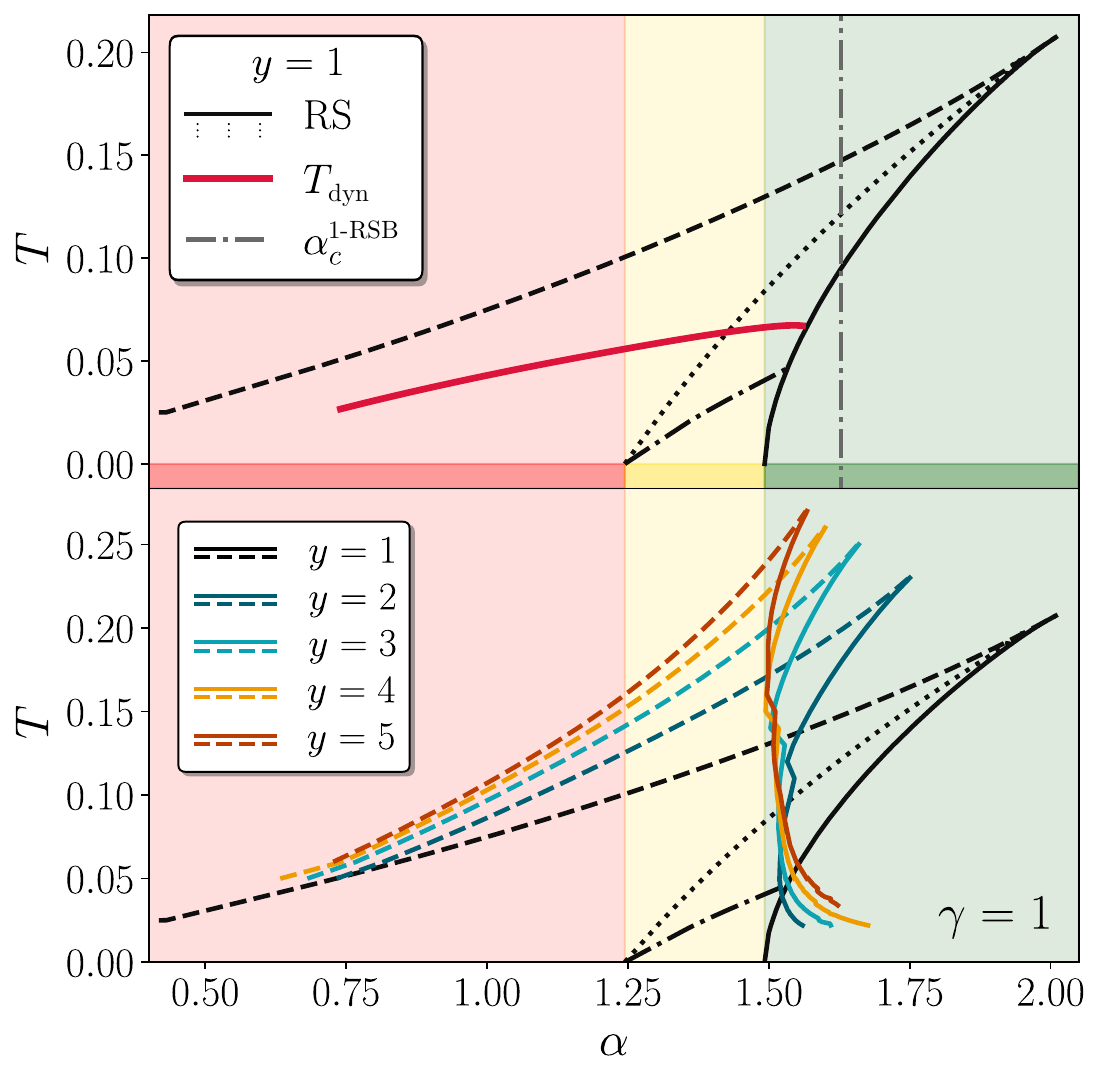}
\put(45,80){{\region{a}}}
\put(51,66.5){{\region{b}}}
\put(65.5,62.5){{\region{c}}}
\put(63.5,55.5){{\region{d}}}
\put(82,68.0){{\region{e}}}
\end{overpic}
\caption{$\left(\alpha,T\right)$-Equilibrium phase diagram of the single, non-replicated, binary perceptron \textbf{(top panel)} and the replicated perceptron according to Eq.~\eqref{eq:HamiltonianCoupledPerceptrons} for a fixed coupling $\gamma\!=\!1$ and $y$ multiple replicas \textbf{(bottom panel)}. The color shades refer respectively to the \textit{impossible} (red), \textit{hard} (yellow), \textit{easy} (green) inference phases as extrapolated from $T\!=\!0$. The \textbf{(a)}-\textbf{(e)} regions are discussed in the text.\label{fig:DP}}
\end{figure}
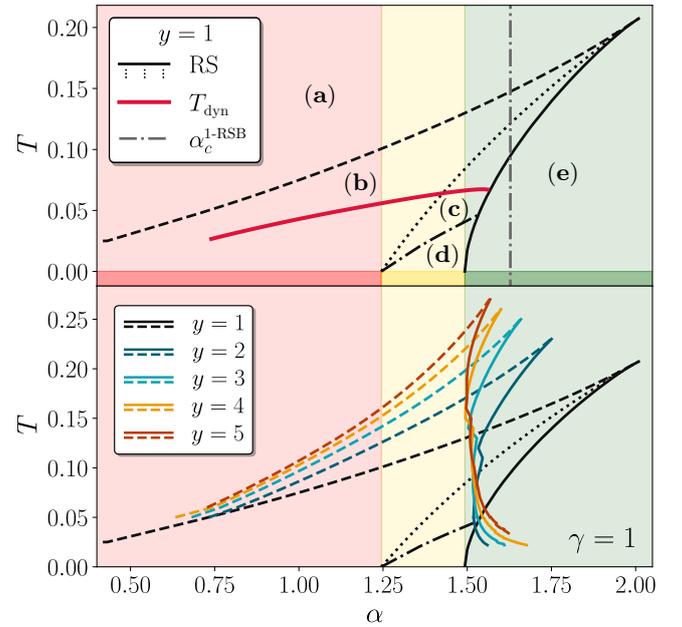

\noindent \textit{Results --} 
It is instructive to first discuss the phase diagram of the single perceptron (i.e. $y\!=\!1$) under the RS ansatz, shown in Fig.~\ref{fig:DP}--top, and previously derived in~\cite{horner_dynamics_1992}.
For $0\!<\! T\!\lesssim \!0.2$ five different equilibrium regimes are found as the fraction of training examples $\alpha$ is increased. Above the dashed line (region \region{a}), the free entropy has a unique maximum with $R\! <\! 1$, which corresponds to a solution with imperfect \textit{generalization}.
Below the dashed line {(region \region{b})}, the teacher's solution  ($R\!\simeq\!1$) appears as a metastable (i.e. subdominant in the free entropy) fixed point, and it becomes dominant after crossing the dotted line (region \region{c}). Finally, beyond the spinodal solid line, in \region{e}, the poor generalization fixed point disappears and only the teacher solution remains. The region \region{d} (below the dash-dotted line) corresponds instead to the \textit{spin-glass} phase where the solution with poor-generalization ($R < 1$) has negative RS entropy. This means that a replica symmetry breaking ansatz (RSB, with $1$ or more steps) would be required to correctly capture the model's thermodynamics. 
The colored areas in Fig.~\ref{fig:DP} stand for the three inference phases present in the $T\!=\!0$ phase diagram~\citep{zdeborova2016statistical}: the \textit{impossible} (red) phase for $\alpha <\alpha_{\text{IT}}\approx 1.24$, where the teacher solution is surrounded by an exponential number of ground states with $0$ training error and therefore it is subdominant in terms of entropy~\citep{baldassisubdominant2015}; the \textit{hard} (yellow) phase $\alpha_{\text{IT}} \!<\! \alpha \!<\!\alpha_{\text{c}}\!\approx\! 1.49$ where although the teacher is now dominant (after the 1st order transition at $\alpha_{\text{IT}}$) there still exists a sub-extensive (in $N$) amount of ground-states with 0 training error, with a disconnected 1RSB like structure; finally, the
\textit{easy} (green) phase for $\alpha > \alpha_{\text{c}}$ where according to the RS ansatz only the teacher configuration remains as a stable maximum of the free-entropy. At $T=0$, the hard-easy boundary $\alpha_{\text{c}}$ also corresponds to the algorithmic threshold for the Approximate Message Passing (AMP) algorithm (a numerical check on finite-size systems is discussed in Appendix \ref{app:AMP}).\\
From the phase diagram in Fig.~\ref{fig:DP}--top, one should be able in principle to infer the performance of a Simulated Annealing (SA) experiment (or, equivalently, a learning process modeled by a slow decrease in the allowable number of errors) for a given value of $\alpha$. In particular, for $\alpha \leq \alpha_{\text{c}}$, SA (or AMP, or any other known polynomial algorithm) is not be able to find the teacher solution, while for $\alpha > \alpha_{\text{c}}$, SA should in principle easily find the planted solution because the equilibrium measure is dominated by configurations close to the teacher vector. However, as pointed out in the next section, SA gets trapped into spurious minima of the energy with non-zero training (and generalization) error even in a finite interval in the easy phase $\alpha > \alpha_c$. At non-zero temperature, the presence of spurious minima can be tracked by computing the dynamic transition temperature $T_{\text{dyn}}$ \citep{franz:jpa-00247146,monasson1995structural}, shown in red in Fig.~\ref{fig:DP}-top: this is the temperature at which spurious local minima with a 1-RSB structure appear, that can act as algorithmic traps for SA. Additional details about this quantity and how to compute it are given in Appendix~\ref{app:Tdyn}, obtained using the procedure discussed in Refs.~\citep{10.21468/SciPostPhys.2.3.019,monasson1995structural}. Following the reason of~\citep{angelini_limits_2023}, SA should always be trapped into spurious minima of the energy as long as the dynamic transition (red line) is met before the spinodal of the imperfect generalization solution (solid black line in Fig.\ref{fig:DP}) upon decreasing the temperature; therefore, the intersection between the two should correspond to the algorithmic threshold for SA, so that $\alpha_c^{\text{SA}}\approx 1.56$. In particular, for $\alpha > \alpha^{\text{SA}}$, SA should always find the solution by melting towards the planted configuration when the spinodal (solid black line in Fig.~\ref{fig:DP}-top) is crossed. For the sake of completeness, we recall that another 1RSB-like analysis was performed in~\citep{sompolinskylearningfromexamples} in the $T\!=\!0$ model: the authors revealed the presence of $1$-RSB frozen metastable states with poor generalization, which cease to exist at $\alpha_{\text{c}}^{\text{RSB}}\!\approx\!1.628$. The presence of such 1RSB states at $T=0$ does not affect the capabilities Bayes-optimal algorithms such as AMP in the regime $\alpha\!>\!\alpha_{c}$ as discussed e.g. in Ref.~\citep{GlassyNatureSilvioSparserank1}: indeed, the algorithmic threshold of AMP is exactly $\alpha_c\approx 1.49$. Conversely, a thermal algorithm like SA is severely affected by these metastable states, although the interplay between the threshold computed in~\citep{sompolinskylearningfromexamples} and the dynamic transition temperature is not yet clear.
Apart on this difference between the 1RSB analysis which is not yet clarified, the general conclusion is that that the higher $\alpha \!>\! \alpha^{\text{SA}}$ is, the easier it should be to find the teacher, as the free entropy landscape becomes smoother as more training examples are available: in particular, on the right side of the intersection between the two spinodals and the transition line (i.e. $\alpha \gtrsim 2.0$), there is no phase transition at all and the generalization error becomes a smooth decreasing function of temperature: in this regime, the energy landscape becomes trivial and the system behaves as an effective paramagnet with a strong external field pointing towards the planted configuration.

Fig.~\ref{fig:DP}-bottom illustrates how the spinodal lines change when $y$ coupled replicas are considered (in this case for $\gamma\!=\!1$). The most interesting phenomenon is that the right spinodal line, which marks the transition to perfect generalization, is bending towards lower values of $\alpha$. Therefore, in the range of $\alpha\!>\!\alpha_{\text{c}}$ the free entropy landscape becomes smoother just by increasing the number of coupled replicas. 
 In terms of performing a temperature annealing, this means that the student perceptrons should not encounter the poorly generalizing glassy states, as they would melt to the teacher solution before the dynamic transition. In practice, this phase diagram shows us that many coupled students need to go through fewer data examples to perfectly deduce the teacher's rule. The effect of the coupling $\gamma$ in the phase diagram is discussed in Appendix~\ref{app:effect_gamma}. A more detailed analysis on the disappearance of metastable glassy states in the coupled system would require a $1$-RSB analysis, which is left for future investigations.

\noindent \textit{Numerical experiments --} We now  investigate numerically the effects of the shifting of the critical lines associated with the coupling between students when training a binary perceptron model~\eqref{eq:HamiltonianCoupledPerceptrons} via SA with a fixed cooling rate. Our SA training protocol is constructed as follows: We initialize student weights at high temperature. At each training update, we perform a Monte Carlo sweep (an update of the entire system in random order) and reduce the temperature by $\eta$, keeping $\alpha$ fixed. At each $\alpha$, we repeat the same process $100$ times, using different teacher perceptrons to average out the disorder fluctuations (additional implementation details are given in  Appendix~\ref{app:SAdetails}).
Fig.~\ref{fig:res_RSA}-\region{a} shows the results for a very slow cooling rate $\eta=10^{-5}$ on different number of replicas $y$.
It is striking that the single perceptron (black dots) is not able to find the teacher solution up to $\alpha\approx1.6$, i.e. deep into the \textit{easy} phase: this is in agreement with the presence of metastable states, as found in the 1-RSB analysis, which trap the annealing dynamics and prevent the system from melting towards the teacher solution. 
A numerical determination of the precise algorithmic threshold for SA on the single system is, however, out of reach for computational reasons: first,  because the system is dense, so that a full MC swipe requires $O\left(\alpha N^2\right)$ operations; secondly, increasing $N$ in order to perform a finite-size scaling for the estimation of $\alpha_c^{\text{SA}}$ would require to rescale the annealing rate with a certain power law of $N$ \citep{angelini_limits_2023}, further increasing the computational cost.
The addition of replicas instead leads to a net shift in the probability of finding the planted solution, which appears to saturate very quickly with $y$ and to approach the theoretical threshold $\alpha_{\text{c}}$. Such a performance difference between SA and RSA (i.e. SA on the replicated system) is even more remarkable at much faster annealing rates $\eta$ (see Fig.~\ref{fig:RSA_app}-\ref{fig:RSA_app_effect-eta} in Appendix~\ref{app:SAdetails}), where systems with larger $y$ find the solution faster. We also observe that even when RSA does not find the planted configuration (e.g., in the inference-\textit{hard} phase), it still finds more optimal solutions (in terms of generalization error, see Fig.~\ref{fig:res_RSA}-\region{b}) compared to the non-replicated system, as previously noted in~\citep{baldassisubdominant2015}.
The change in the flattering of the landscape with $y$ is particularly clear in Fig.~\ref{fig:res_RSA} (bottom panel): the RSA trajectories show increasingly smoother behavior (in terms of generalization error) as the number of students increases, but at the same value of $\alpha$. Note also the good agreement between simulations and theory (in white outlined lines); the disagreement between theory and simulations for $y=1$ and $\alpha=1.56$ (Fig.~\ref{fig:res_RSA}-\region{c}) is expected to be due to the failure of the theory at the RS level.\\
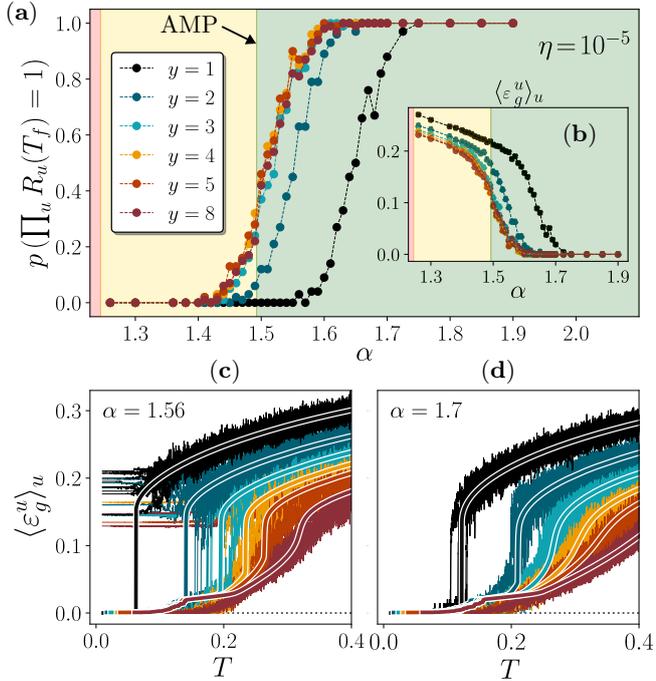
\begin{figure}[t!]
    \centering
    \begin{overpic}[trim=0 10 0 5,clip,width=1.\columnwidth]{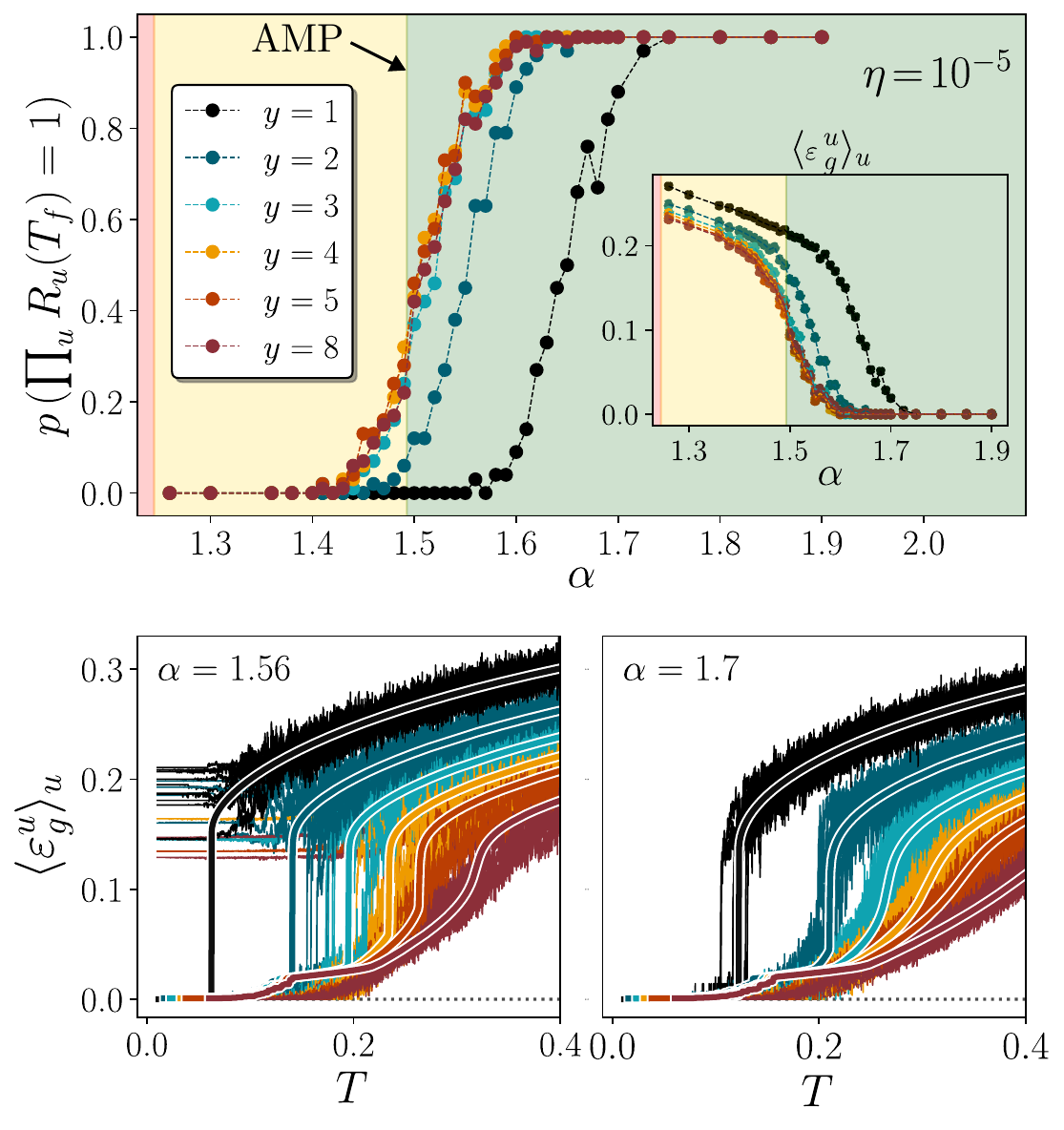}
\put(0,98){{\region{a}}}
\put(82,80){{\region{b}}}
\put(30,45){{\region{c}}}
\put(70,45){{\region{d}}}
\end{overpic}
    \caption{Numerical performances of replicated SA (RSA) with an annealing rate $\eta=10^{-5}$. In black lines we show the performance of the single perceptron and in colors that of the coupled perceptron with different $y$. \textbf{(a)}: empirical probability (from $100$ training instances) of finding the teacher configuration at the end of the annealing process, as a function of $\alpha$, for $N\!=\!2001$, $\gamma\!=\!1$. The inset \textbf{(b)} shows the corresponding mean generalization error $\varepsilon_g = \pi^{-1}\text{acos}(R)$. The colored regions are the same as in Fig.~\ref{fig:DP}: in particular, the hard-easy phase boundary corresponds to the algorithmic threshold for AMP at $T=0$. \textbf{(c)} and \textbf{(d)}: Examples of typical annealing trajectories for $2$ values of $\alpha$ ($10$ training trajectories are shown for each $y$). The settings and color coding are the same as in the top panel. The white outlined lines show the analytical result obtained from solving the RS self-consistent equations, starting from a poorly generalizing solution ($R\!\ll\! 1$) at $T\!=\!0.5$, and following the fixed point as the temperature is linearly decreased.}
    \label{fig:res_RSA}
\end{figure}
\textit{Discussion --} In this paper, we have illustrated how the phase diagram of the binary perceptron in the teacher-student scenario changes as more and more students try to learn the same teacher rule while being coupled to each other. From this phase diagram, we can draw conclusions not only about the number of examples required for perfect recovery (as a function of the number of collaborating students), which is much smaller than for a single perceptron, but also about the performance of a thermal learning procedures such as SA, or eventually stochastic gradient descent (on models with continuous weights).

The coupled perceptron model discussed here provides a toy model to shed light on the effectiveness of students' collective learning in acquiring a particular set of predetermined rules, as exploited in recent real-life experiences (see, e.g.~\citep{kampal}).
Within the framework of this rudimentary model, further generalizations could explore the question of whether diversifying the examples for each student improves the learning experience or whether alternative learning paths could accelerate learning. In particular, our approach could be used to rationalize the impact of federated or collaborative learning in machine learning, a decentralized approach to training models~\citep{yang2019federated}. In both contexts, our phase diagrams facilitate the determination of the ideal number of collaborating teams, the best number of examples or the appropriate learning pace to ensure optimal learning.\\ From a theoretical point of view, we believe that this work contributes to a better understanding of the role of the robust ensemble, which has been recently proposed to develop novel algorithmic schemes for solving constraint satisfaction problems. In particular, our results seem to confirm that RSA is a generic and robust inference algorithm that has a close-to-Bayes optimal threshold for a sufficient number of coupled replicas, as recently argued in another setup~\citep{angelini_limits_2023}. Our results also suggest that coupled neural networks can in principle perform better in terms of generalization error for a fixed amount of data. It would be interesting to verify this effect on more complex architectures for a supervised learning task or on other examples of planted models with a hard inference phase. In this regard, a more analytical study of the bending of the phase diagram and the resulting change in the nature of the phase transition in the inference \textit{easy} phase in the limit $y\to\infty$ will be the subject of a follow-up work. \\ 
\begin{acknowledgments}
We thank Carlo Lucibello, Federico Ricci-Tersenghi, Valentina Ros and Pierfrancesco Urbani for useful discussions and suggestions regarding the article. We further thank Carlo Lucibello for sharing an AMP implementation which we used for comparison. The authors acknowledge financial support by the Comunidad de
Madrid and the Complutense University of Madrid (UCM) through the Atracción de Talento programs (Refs. 2019-T1/TIC-13298 and 2019-T1/TIC-12776),
the Banco Santander and the UCM (grant PR44/21-29937), and Ministerio
de Econom\'{\i}a y Competitividad, Agencia Estatal de Investigaci\'on
and Fondo Europeo de Desarrollo Regional (Ref. PID2021-125506NA-I00) 
MICIU/AEI
/10.13039/501100011033 y por
FEDER, UE
\end{acknowledgments}

\bibliographystyle{apsrev4-2}

\appendix
\onecolumngrid
\newpage

\setcounter{figure}{0}  
\renewcommand{\thefigure}{S\arabic{figure}}

\section{Derivation of Quenched free entropy\label{app:MFTheory}}

In this section we discuss how to compute the averaged quenched free
energy for the model \eqref{eq:HamiltonianCoupledPerceptrons}. As
standard in spin-glass models with quenched disorder, the first
step requires the introduction of a number $n$
of replicas, whose limit $n\to0$ is taken afterwards. These replicas
differ from the ones in the original Hamiltonian because they are independent
and no explicit coupling is present between them. In the following
derivation, and in order to avoid confusion between the two sets of
replicas, we always use indices $a,b\in\left\{ 1,\ldots,n\right\} $
to denote ``fake'' replicas and $u,v\in\left\{ 1,\ldots y\right\} $ to
index students (i.e. real replicas in the original Hamiltonian). Replicating
all the degrees of freedom $w_{i,u}$ $n$ times (with integer $n$)
we can write the replicated partition function as

\begin{equation}
Z^{n}=\sum_{\boldsymbol{w}_{u}^{\left(a\right)}}\exp\left[-\beta\sum_{a,u,\mu}V\left(\Delta_{u,a}^{\mu}\right)+\frac{\gamma}{y}\sum_{a}\sum_{u<v}\boldsymbol{w}_{u}^{\left(a\right)}\cdot\boldsymbol{w}_{v}^{\left(a\right)}\right].\label{eq:Zreplicated}
\end{equation}
 We start by introducing the definition of the stabilities
\begin{align}
\Delta_{u,a}^{\mu} & =\sigma_{0}^{\mu}\frac{\boldsymbol{w}_{u}^{\left(a\right)}\cdot\boldsymbol{\xi}^{\mu}}{\sqrt{N}},\label{eq:stability_def_u_a_mu}
\end{align}
with $\sigma_{0}^{\mu}$ given by Eq. (\ref{eq:def_output_teacher}) in
the main text. Enforcing definitions \eqref{eq:stability_def_u_a_mu} and \eqref{eq:def_output_teacher}
by using delta functions and exploiting their Fourier representation, the partition function (\ref{eq:Zreplicated}) can be re-written as 
\begin{align}
Z^{n} & =\sum_{\boldsymbol{w}_{u}^{\left(a\right)}}\int\prod_{\mu}d\omega_{0}^{\mu}d\hat{\omega}_{0}^{\mu}\int\prod_{\mu,a,u}d\Delta_{u,a}^{\mu}d\hat{\Delta}_{u,a}^{\mu}\exp\left[-\beta\sum_{a,u,\mu}V\left(\Delta_{u,a}^{\mu}\right)+\frac{\gamma}{y}\sum_{a}\sum_{u<v}\boldsymbol{w}_{u}^{\left(a\right)}\cdot\boldsymbol{w}_{v}^{\left(a\right)}\right]\nonumber \\
 & \times\exp\left[\text{i}\sum_{\mu}\omega_{0}^{\mu}\hat{\omega}_{0}^{\mu}-\frac{\text{i}}{\sqrt{N}}\sum_{\mu}\hat{\omega}_{0}^{\mu}\left(\boldsymbol{w}^{0}\cdot\boldsymbol{\xi}^{\mu}\right)+\text{i}\sum_{a,\mu,u}\Delta_{u,a}^{\mu}\hat{\Delta}_{u,a}^{\mu}-\frac{\text{i}}{\sqrt{N}}\sum_{a,\mu,u}\hat{\Delta}_{u,a}^{\mu}\left(\boldsymbol{w}_{u}^{\left(a\right) }\cdot\boldsymbol{\xi}^{\mu}\right) \sigma_{0}^{\mu}\left( \omega_0^{\mu} \right) \right].\label{eq:Znafterintroducing_x0_ya_allmu}
\end{align}
where $\hat{\omega}_{0}^{\mu}$ (resp. $\hat{\Delta}_{u,a}^{\mu}$) are the conjugate variables of $\omega_{0}^{\mu}$ (resp. $\Delta_{u,a}^{\mu}$) introduced through a Fourier transform. The dependency of the teacher label on its input simply follows from the Perceptron classification rule, i.e. $\sigma_0^{\mu}\left(\omega_0^{\mu}\right)=\text{sign} \left(\omega_0^{\mu}\right)$, although in the rest of the calculation we will drop this dependency for notation convenience. It is now easy to perform the average over the disorder given by the pattern components. As for now we do not perform the average over the teacher weight vector, but anyways it will become trivial in the final expression. As specified in the main text, we assume the pattern components to be i.i.d. with binary entries, so that $\xi_{i}^{\mu}\in\left\{ -1,1\right\} $ with equal probability.  The average concerns only the second and fourth
term in the second line of \eqref{eq:Znafterintroducing_x0_ya_allmu}:
\begin{align}
\left\langle e^{-\frac{\text{i}}{\sqrt{N}}\sum_{\mu}\hat{\omega}_{0}^{\mu}\left(\boldsymbol{w}^{0}\cdot\boldsymbol{\xi}^{\mu}\right)-\frac{\text{i}}{\sqrt{N}}\sum_{a,\mu,u}\hat{\Delta}_{u,a}^{\mu}\sigma_{0}^{\mu}\left(\boldsymbol{w}_{u}^{\left(a\right)}\cdot\boldsymbol{\xi}^{\mu}\right)}\right\rangle _{\left\{ \boldsymbol{\xi}^{\mu}\right\} _{\mu=1}^{M}} & =\prod_{i,\mu}\left\langle \exp\left[-\frac{\text{i}}{\sqrt{N}}\xi_{i}^{\mu}\left(\hat{\omega}_{0}^{\mu}w_{i}^{0}+\sigma_{0}^{\mu}\sum_{a,u}\hat{\Delta}_{u,a}^{\mu}w_{i,u}^{\left(a\right)}\right)\right]\right\rangle _{\xi_{i}^{\mu}}\nonumber \\
 & =\prod_{i,\mu}2\text{cosh}\left[\frac{\text{i}}{\sqrt{N}}\left(\hat{\omega}_{0}^{\mu}w_{i}^{0}+\sigma_{0}^{\mu}\sum_{a,u}\hat{\Delta}_{u,a}^{\mu}w_{i,u}^{\left(a\right)}\right)\right]\nonumber \\
 & \approx\exp\left[-\frac{1}{2N}\sum_{i\mu}\left(\hat{\omega}_{0}^{\mu}w_{i}^{0}+\sigma_{0}^{\mu}\sum_{a,u}\hat{\Delta}_{u,a}^{\mu}w_{i,u}^{\left(a\right)}\right)^{2}\right],\label{eq:result_disorderterm}
\end{align}
where in the first line we use the fact that pattern components are
i.i.d. and in the last line we expanded for $N\to\infty$, keeping only the first
order, the other ones being subdominant in the thermodynamic
limit. 

As usual, the disorder average results into an effective coupling
between replicas $a,b$, and another coupling will also be taken into
account between students in the same replica $a$ on top of the explicit
one in the Hamiltonian. We can now introduce a set of order parameters,
namely the overlap between student $u$ (in replica $a$) with the
teacher and the two-replica overlap between two student vectors $u,v$,
respectively given by:
\begin{align}
R_{a}^{u} & =\frac{1}{N}\sum_{i}w_{i,u}^{\left(a\right)}w_{i}^{0}\label{eq:R_au_def_app}\\
q_{ab}^{uv} & =\frac{1}{N}\sum_{i}w_{i,u}^{\left(a\right)}w_{i,v}^{\left(b\right)}\label{eq:q_ab_uv_def_app}
\end{align}
It is easy to visualize the overlap matrix \eqref{eq:q_ab_uv_def_app} in a block-matrix
form. We can indeed write a $n\times n$ block matrix of the type
\begin{equation}
\boldsymbol{\mathcal{Q}}=\begin{bmatrix}\boldsymbol{\mathcal{Q}}_{11} & \dots & \boldsymbol{\mathcal{Q}}_{1n}\\
\vdots & \ddots & \vdots\\
\boldsymbol{\mathcal{Q}}_{n1} & \dots & \boldsymbol{\mathcal{Q}}_{nn}
\end{bmatrix}\label{eq:overlapmatrix_Blocks}
\end{equation} where each inner matrix $\boldsymbol{\mathcal{Q}}_{ab}$ has dimension $y\times y$. Each of these matrices describes the typical 2-point overlap between students with two generic replica indices $a$ and $b$. Exploiting trivial
symmetries under indices permutations, the number of independent overlaps is be equal to $n\binom{y}{2}+\binom{n}{2}y^{2}=\binom{ny}{2}$.
Substituting \eqref{eq:result_disorderterm} and enforcing definitions \eqref{eq:R_au_def_app}-\eqref{eq:q_ab_uv_def_app}
through delta functions, we can rewrite the averaged partition function as
\begin{multline}
\left\langle Z^{n}\right\rangle_{\left\{ \boldsymbol{\xi}^{\mu}\right\} _{\mu=1}^{M}} =\sum_{\boldsymbol{w}_{u}^{\left(a\right)}}\int\prod_{a,u}dR_{a}^{u}d\hat{R}_{a}^{u}\int\prod_{a,b,u,v\in\mathcal{P}}dq_{ab}d\hat{q}_{ab}\int\prod_{\mu}d\omega_{0}^{\mu}d\hat{\omega}_{0}^{\mu}\int\prod_{\mu,a,u}\prod_{\mu,a,u}d\Delta_{u,a}^{\mu}d\hat{\Delta}_{u,a}^{\mu}\\
\exp\left[-\beta\sum_{a,u,\mu}V\left(\Delta_{u,a}^{\mu}\right)+\text{i}\sum_{\mu}\omega_{0}^{\mu}\hat{\omega}_{0}^{\mu}+\text{i}\sum_{a,\mu,u}\Delta_{u,a}^{\mu}\hat{\Delta}_{u,a}^{\mu}+\text{i}N\sum_{a,u}R_{a}^{u}\hat{R}_{a}^{u}+\text{i}N\sum_{a,b,u,v\in\mathcal{P}}q_{ab}^{uv}\hat{q}_{ab}^{uv}-\frac{1}{2}\sum_{\mu}\left(\hat{\omega}_{0}^{\mu}\right)^{2} \right.\\
-\sum_{\mu}\hat{\omega}_{0}^{\mu}\sigma_{0}^{\mu}\sum_{a,u}\hat{\Delta}_{u,a}^{\mu}R_{a}^{u}-\frac{1}{2}\sum_{\mu,a,u}\left(\hat{\Delta}_{u,a}^{\mu}\right)^{2}-\sum_{\mu,a}\sum_{u<v}q_{aa}^{uv}\hat{\Delta}_{u,a}^{\mu}\hat{\Delta}_{v,a}^{\mu}-\sum_{\mu}\sum_{a<b}\sum_{u,v}q_{ab}^{uv}\hat{\Delta}_{u,a}^{\mu}\hat{\Delta}_{v,b}^{\mu}+\\
\left.+\frac{\gamma}{y}\sum_{a}\sum_{u<v}\boldsymbol{w}_{u}^{\left(a\right)}\cdot\boldsymbol{w}_{v}^{\left(a\right)}-\text{i}\sum_{a,u}\hat{R}_{a}^{u}\sum_{i}w_{i,u}^{\left(a\right)}w_{i}^{0}-\text{i}\sum_{a,b,u,v\in\mathcal{P}}\hat{q}_{ab}^{uv}\sum_{i}w_{i,u}^{\left(a\right)}w_{i,v}^{\left(b\right)}\right]\label{eq:Zn_afterintroducingoverlaps}
\end{multline}
where the symbol $\mathcal{P}$ is a short-hand notation to indicate
all the possible independent overlaps. In particular, $\sum_{a,b,u,v\in\mathcal{P}}=\sum_{a}\sum_{u<v}+\sum_{a<b}\sum_{u,v}$.
We can notice now how the integrals over $\mu$-dependent quantities
can be factorized, as well as the sum over weight components $i$.
We can therefore rewrite Eq. (\ref{eq:Zn_afterintroducingoverlaps})
in a saddle point form. Using the property that the teacher vector
components are i.i.d, we can write

\begin{figure}[t]
\includegraphics[width=0.6\columnwidth]{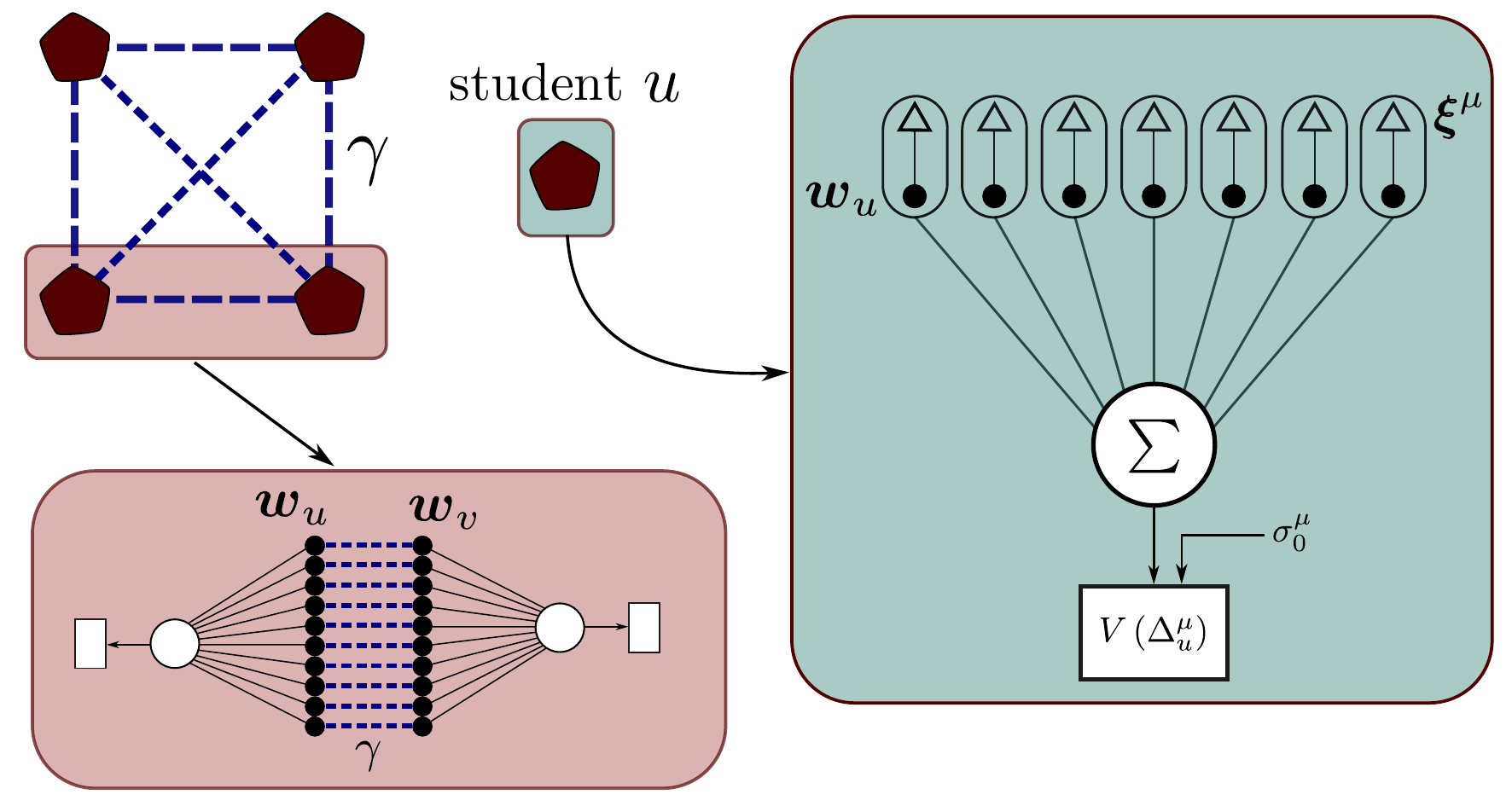}
\caption{Schematic representation of the model defined by Eq. ~(\ref{eq:HamiltonianCoupledPerceptrons}) for  $y=4$ student perceptrons interacting over a fully connected graph. The dashed blue lines represent the coupling $\gamma$ between students.}\label{fig:sketch}
\end{figure}

\begin{equation}
\left\langle Z^{n}\right\rangle _{\left\{ \boldsymbol{\xi}^{\mu}\right\} _{\mu=1}^{M},\boldsymbol{w}_{0}}=\int\prod_{a,u}dR_{a}^{u}d\hat{R}_{a}^{u}\int\prod_{a,b,u,v\in\mathcal{P}}dq_{ab}d\hat{q}_{ab}\exp\left[Ny\mathcal{G}\left(\left\{ R_{a}^{u},q_{ab}^{uv},\hat{R}_{a}^{u},\hat{q}_{ab}^{uv}\right\} \right)\right]\label{eq:Zn_saddlepoint}
\end{equation}

\begin{equation}
\mathcal{G}\left(\left\{ R_{a}^{u},q_{ab}^{uv},\hat{R}_{a}^{u},\hat{q}_{ab}^{uv}\right\} \right)=\frac{\text{i}}{y}\sum_{a,u}R_{a}^{u}\hat{R}_{a}^{u}+\frac{\text{i}}{y}\sum_{a,b,u,v\in\mathcal{P}}q_{ab}^{uv}\hat{q}_{ab}^{uv}+\frac{\alpha}{y}G_{E}\left(\left\{ R_{a}^{u},q_{ab}^{uv}\right\} \right)+\frac{1}{y}G_{I}\left(\left\{ \hat{R}_{a}^{u},\hat{q}_{ab}^{uv}\right\} \right)\label{eq:mf_free_energy_generic}
\end{equation}
where $\mathcal{G}$ plays the role of a free entropy with opposite
sign. The equilibrium behavior is thus determined by the maximum of Eq. \eqref{eq:mf_free_energy_generic} w.r.t. the order parameters. The quantities and $G_{E}$ and $G_{I}$ represent the usual entropic and energetic terms, respectively given by
\begin{align}
G_{I} & =\mathbb{E}_{w^{0}}\log\sum_{w_{u}^{a}\in\left\{ -1,1\right\} }\exp\left[\frac{\gamma}{y}\sum_{a}\sum_{u<v}w_{u}^{a}w_{v}^{a}-\text{i}w^{0}\sum_{a,u}\hat{R}_{a}^{u}w_{u}^{a}-\text{i}\sum_{a,b,u,v\in\mathcal{P}}\hat{q}_{ab}^{uv}w_{u}^{a}w_{v}^{b}\right]\label{eq:GI_generic_binaryweights}
\end{align}

\begin{multline}
G_{E}=\log\int d\omega_{0}d\hat{\omega}_{0}d\Delta_{u}^{a}d\hat{\Delta}_{u}^{a}\exp\left[-\beta\sum_{a,u}V\left(\Delta_{u}^{a}\right)+\text{i}\omega_{0}\hat{\omega}_{0}+\text{i}\sum_{a,u}\Delta_{u}^{a}\hat{\Delta}_{u}^{a}-\frac{1}{2}\hat{\omega}_{0}^{2}-\frac{1}{2}\sum_{a,u}\left(\hat{\Delta}_{u,a}\right)^{2}+\right.\\
\left.-\hat{\omega}_{0}\sigma_{0}\sum_{a,u}\hat{\Delta}_{u}^{a}R_{a}^{u}-\frac{1}{2}\sum_{a,u\neq v}\hat{\Delta}_{u}^{a}\hat{\Delta}_{v}^{a}q_{aa}^{uv}-\frac{1}{2}\sum_{a\neq b}\sum_{u,v}\hat{\Delta}_{u}^{a}\hat{\Delta}_{v}^{b}q_{ab}^{uv}\right]\label{eq:GE_energeticterm_generic}
\end{multline}
where in the last line we exploited the fact that the teacher vector
components are i.i.d. by assumption. Notice that the expectation over one representative component $w_{0}$ becomes dummy by means of
a gauge transformation $w_{u}^{a}\to w^{0}w_{u}^{a}$ for
all the weight components, so we will drop it from now on. Before going on, note that the integral over
$\hat{\omega}_{0}$ in Eq. (\ref{eq:GE_energeticterm_generic}) can
be carried out explicitly, leading to

\begin{multline}
G_{E}=\log\int\mathcal{D}\omega_{0}d\Delta_{u}^{a}d\hat{\Delta}_{u}^{a}\exp\left\{-\beta\sum_{a,u}V\left(\Delta_{u}^{a}\right)+\text{i}\sum_{a,u}\hat{\Delta}_{u}^{a}\left(\Delta_{u}^{a}-\omega_{0}\sigma_{0}R_{a}^{u}\right)-\frac{1}{2}\sum_{a,u}\left[1-\left(R_{a}^{u}\right)^{2}\right]\left(\hat{\Delta}_{u,a}^{\mu}\right)^{2}\right.\\
\left.-\frac{1}{2}\sum_{a,u\neq v}\hat{\Delta}_{u}^{a}\hat{\Delta}_{v}^{a}\left(q_{aa}^{uv}-R_{a}^{u}R_{a}^{v}\right)-\frac{1}{2}\sum_{a\neq b}\sum_{u,v}\hat{\Delta}_{u}^{a}\hat{\Delta}_{v}^{b}\left(q_{ab}^{uv}-R_{a}^{u}R_{b}^{v}\right)\right\}\label{eq:GE_afterIntegralomega0hat}
\end{multline}

where $\mathcal{D}x=e^{-x^{2}\slash2}dx\slash\sqrt{2\pi}$ denotes
the standard Gaussian probability measure.

\subsection{Replica symmetric ansatz on both spaces\label{subsec:Ansatz_RRS_nodiag}}
The simplest possible ansatz corresponds to assume a permutation symmetry over \textit{both} replica spaces. For what concerns the "fake"-replicated space, this is the simplest choice in any spin glass model \citep{mvpbook}. On the other hand, a symmetry ansatz between the pairwise correlations between students (i.e. the overlaps $q_{aa}^{uv}$) comes naturally from the fully-connected topology of interactions between the students' weight vectors as assumed in Eq.~\eqref{eq:HamiltonianCoupledPerceptrons} in the main text. Choosing different topologies would imply to parametrize the matrices $\boldsymbol{\mathcal{Q}}_{aa}$ in such a way to reflect the behavior of non-connected correlations in that specific graph, which is itself a non-trivial problem unless in very specific architectures (e.g. trees or planar graphs). However, restricting to the fully-connected topology as in the main text, within this extended RS assumption we have just only 3 order parameters (and their conjugates) left, namely:
\begin{subequations}
\begin{alignat}{3}
R_{a}^{u} & =R\quad\forall u,a & \qquad\qquad & q_{aa}^{uv}=p\quad\forall u\neq v,a & \qquad\qquad & q_{ab}^{uv}=q\quad\forall u,v,a\neq b\\
\hat{R}_{a}^{u} & =\text{i}\hat{R}\quad\forall u,a &  & \hat{q}_{aa}^{uv}=\text{i}\hat{p}\quad\forall u\neq v,a &  & \hat{q}_{ab}^{uv}=\text{i}\hat{q}\quad\forall u,v,a\neq b
\end{alignat}
\label{subeq:RS_ansatz_RRS_nodiag}
\end{subequations}
By inserting the above ansatz into Eqs. \eqref{eq:mf_free_energy_generic}-\eqref{eq:GE_afterIntegralomega0hat}-\eqref{eq:GI_generic_binaryweights} and taking the $n\to 0 $ limit, after some calculations we can rewrite the RS quenched free entropy as:

\begin{equation}
    \mathcal{G} =-R\hat{R}-\frac{\left(y-1\right)}{2}p\hat{p}+\frac{\hat{q}}{2}\left(yq-1\right)+\frac{2\alpha}{y}\int\mathcal{D}t\,\Phi\left(at\right)\log\int\mathcal{D}\tau\Xi^{y}+\frac{1}{y}\int\mathcal{D}z\log \mathcal{Z}_{\left(y\right)},\label{eq:free_en_RRS_nodiag}
\end{equation}
where \begin{align}
\Xi & =\int\frac{d\Delta}{\sqrt{1-p}}\exp\left[-\beta V\left(\Delta\right)-\frac{1}{2\left(1-p\right)}\left(\Delta+\sqrt{q}t+\sqrt{p-q}\tau\right)^{2}\right],\label{eq:InnerK_RRS_nodiag}\\
&\mathcal{Z}_{\left(y\right)} =\sum_{w_{u}\in \pm 1}\exp\left[\left(\frac{\gamma}{y}+\hat{p}-\hat{q}\right)\sum_{u<v}w_{u}w_{v}+\left(w^{0}\hat{R}+\sqrt{\hat{q}}z\right)\sum_{u}w_{u}\right],\label{eq:InnerZ_RRS_Nodiag}
\end{align}
with $\Phi(x)=\int_x^{\infty} \mathcal{D}z = \frac{1}{2}\text{erfc}\left(x\slash \sqrt{2}\right)$ and for notation's shortness we defined the following three quantities:
\begin{equation}
a=\frac{R}{\sqrt{q-R^{2}}};\qquad\qquad b=\sqrt{\frac{q}{1-p}};\qquad\qquad c=\sqrt{\frac{p-q}{1-p}}.\label{eq:def_abc}
\end{equation}
Concerning the energetic term in \eqref{eq:free_en_RRS_nodiag}, explicit formulas for $\Xi$ \eqref{eq:InnerK_RRS_nodiag} depend on the specific choice for the potential. For a potential of the type $V\left(\Delta\right) = \left(-\Delta\right)^{\nu} \Theta \left(-\Delta\right) $, with $\nu=1$, it reads: \begin{equation}
V\left(\Delta\right)  \,= \, -\Delta\Theta\left(-\Delta\right)  \quad\longrightarrow\quad \Xi = \Phi\left(bt+c\tau\right)+e^{\frac{\beta^{2}\left(1-p\right)}{2}-\beta\sqrt{1-p}\left(bt+c\tau\right)}\Phi\left(\beta\sqrt{1-p}-bt-c\tau\right)\label{eq:innerK_RRS_nodiag_PerceptronRUle}
\end{equation}    

Interestingly, the entropic term can be seen as an averaged free entropy (apart on a sign)
of a reduced system of $y$ degrees of freedom (in this case, binary spins due to the binary nature of the starting weights): speficically, at fixed $z$ Eq. \eqref{eq:InnerZ_RRS_Nodiag} defines the partition function of a Curie Weiss model of $y$ spins with an
effective ferromagnetic coupling and a global external field given by the sum of two terms: a signal one proportional to $\hat{R}$, and a \textit{global} Gaussian field $z$ to be further averaged over.

\subsubsection{Saddle point equations}
The equilibrium behavior of the system at fixed control parameters is given by the maximum of $\mathcal{G}$ w.r.t. all the order parameters, whose values are determined by imposing stationarity of the free entropy. After some calculations, we can write the self-consistent equations for the $6$ order parameters $\left(R,q,p, \hat{R}, \hat{q}, \hat{p}\right)$ as
\begin{subequations}
\begin{align}
\hat{R} & =\sqrt{\frac{2}{\pi}}\alpha\beta\sqrt{\frac{q}{q-R^{2}}}\int\mathcal{D}te^{-\frac{a^{2}t^{2}}{2}}\frac{\int\mathcal{D}\tau\mathcal{T}\Phi\left(\beta\sqrt{1-p}-bt-c\tau\right)\Xi^{y-1}}{\int\mathcal{D}\tau\Xi^{y}},\\
\hat{q} & =2\alpha\beta^{2}\int\mathcal{D}t\,\Phi\left(at\right)\frac{\left[\int\mathcal{D}\tau\Xi^{y-1}\mathcal{T}\Phi\left(\beta\sqrt{1-p}-bt-c\tau\right)\right]^{2}}{\left[\int\mathcal{D}\tau\Xi^{y}\right]^{2}},\\
\hat{p} & =2\alpha\beta^{2}\int \mathcal{D}t\frac{\Phi\left(at\right)}{\int \mathcal{D}\tau\Xi^{y}}\int \mathcal{D}\tau\mathcal{T}^{2}\Xi^{y-2}\Phi^{2}\left(\beta\sqrt{1-p}-bt-c\tau\right),
\end{align}
\label{subeq:SP_equations_coupled_Rpq}
\end{subequations}
 for the conjugate order parameters - where we defined for convenience $\mathcal{T} = e^{\frac{\beta^{2}\left(1-p\right)}{2}-\beta\sqrt{1-p}\left(bt+c\tau\right)}$ -, and 
\begin{subequations}
\begin{align}
R & =\int \mathcal{D}z\,\Omega\left(w_{1}\right), \label{eq:SPeq_R}\\
q & = \int \mathcal{D}z\,\Omega\left(w_{1}\right) \Omega \left(w_{2}\right)= \int \mathcal{D}z\,\Omega^{2}\left(w_{1}\right), \label{eq:SPeq_q}\\
p & =\int \mathcal{D}z\,\Omega\left(w_{1}w_{2}\right), \label{eq:SPeq_p}
\end{align}\label{subeq:SPeq_Rqp}
\end{subequations}
with  $\Omega\left(O\right)$ denoting the expectation of a generic observable $O$ for the partition function \eqref{eq:InnerZ_RRS_Nodiag}, namely:
\begin{equation}
\Omega\left(O\right)=\frac{1}{\mathcal{Z}_{\left(y\right)}\left(\hat{R},\hat{p},\hat{q},\gamma\right)}\sum_{w_{u} \in \pm 1} O \exp\left[\left(\frac{\gamma}{y}+\hat{p}-\hat{q}\right)\sum_{u<v}w_{u}w_{v}+\left(\hat{R}+\sqrt{\hat{q}}z\right)\sum_{u}w_{u}\right].\label{eq:Omega_observable}
\end{equation} 
In the limit $\gamma \to 0$ the different students are un-coupled and the free entropy becomes identical to that of the non-replicated model: in other terms, the only possible solution of the saddle point equations at $\gamma=0$ is $p=q$ and $\hat{p}=\hat{q}$.

\subsection{Effect of $\gamma$ \label{app:effect_gamma}}
\begin{figure}[ht]
    \centering
    \includegraphics[width=0.45\textwidth]{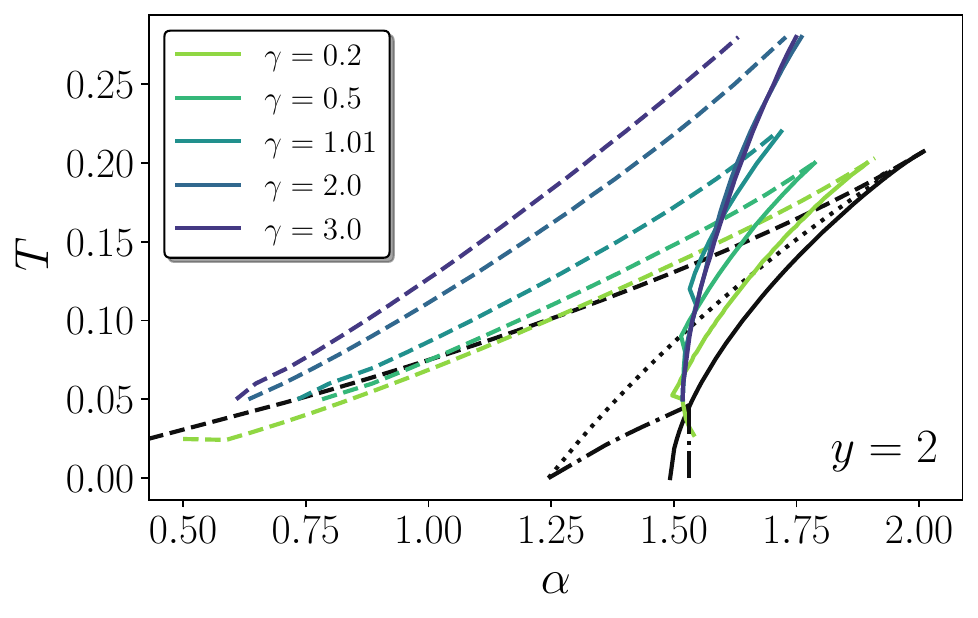}
    \caption{ Phase diagram of $y=2$ coupled binary perceptrons in the $\left(T-\alpha\right)$ plane at different values of $\gamma$ (values given in the legend). Comparison with the single perceptron in the RS approach (the black lines). }
    \label{fig:DP_effectGamma}
\end{figure}

In Fig.~\ref{fig:DP_effectGamma} we show the effect of changing $\gamma$ at a fixed number of coupled students $y$. The spinodal lines continuously approach the result for the single system $y=1$ when $\gamma \to 0$, confirming the validity of the theory. This holds independently on the number $y$ of students. As long as $\gamma$ is increased, the phase diagram seems also to shift at higher temperatures: this would imply that the larger $\gamma$, the higher the starting temperature for RSA should be, in such a way to start the annealing in a regime where the interaction between replicas has a lower effect w.r.t. the cost function of each student.

\section{MF theory of Single Perceptron\label{app:SinglePerceptron}}
For the sake of completeness, in this appendix we report the expressions of the free entropy - and the corresponding self-consistent equations for the order parameters - for the single Perceptron (i.e at $y=1$), obtained both through a replica-symmetric (RS) and 1-step replica-symmetry-breaking (1-RSB) ansatzs. We omit the derivation of all the following expressions as they have been extensively computed in several works (see e.g. \citep{sompolinskylearningfromexamples,statmechlearningruleBiel,engel_statistical_2001} and references therein). In both cases, the $2$ control parameters are simply the fraction $\alpha$ of examples provided to the student and the temperature $T$. The following expressions are reported considering the same learning rule as in the main text, i.e $V\left(\Delta\right)=-\Delta \Theta \left(-\Delta\right)$, where $\Delta$ is the stability parameter.
\subsection{RS free entropy\label{app:RSfree_energy_single}}
In this case, there are two order parameters $R,q$ (and their conjugates $\hat{R},\hat{q}$) where $R$ is the typical overlap between student and teacher's weights, and $q$ is the (unique by assumption) $2$-replica overlap. The free entropy as a function of these order parameters ($R,q,\hat{R},\hat{q}$) and the corresponding self-consistent equations read:
 \begin{align}
\mathcal{G}&=-R\hat{R}+\frac{1}{2}\left(q-1\right)\hat{q}+2\alpha\int\mathcal{D}t\,\Phi \left(at\right)\log \Xi+\int\mathcal{D}z\log2\text{cosh}\left(\hat{R}+\sqrt{\hat{q}}z\right);\\
\Xi &= \Phi \left(bt\right)+e^{\frac{\beta^{2}\left(1-q\right)}{2}-\beta\sqrt{q}t}\Phi \left(\beta\sqrt{1-q}-bt\right);\label{eq:RS_free_energy}
\end{align}
and
\begin{subequations}
\begin{alignat}{2}
R & = \int \mathcal{D}z\, \text{tanh} \left(\hat{R}+\sqrt{\hat{q}}z\right)  ;\qquad\quad & \hat{R} & = \frac{2\alpha\beta}{\sqrt{2\pi}}\int\mathcal{D}z\left[1+\frac{\Phi\left(vz\right)}{\Phi\left(\beta\sqrt{1-q}-vz\right)}e^{\beta\sqrt{q-R^{2}}z-\frac{\beta^{2}\left(1-q\right)}{2}}\right]^{-1};\\
q & = \int \mathcal{D}z \,\text{tanh}^2 \left(\hat{R}+\sqrt{\hat{q}}z\right) \, ; & \hat{q} & =2\alpha\beta^{2}\int\mathcal{D}t\Phi\left(at\right)\left[1+\frac{\Phi\left(bt\right)}{\Phi\left(\beta\sqrt{1-q}-bt\right)}e^{\beta\sqrt{q}t-\frac{\beta^{2}\left(1-q\right)}{2}}\right]^{-2},
\end{alignat}\label{eq:selfconsistentRS}
\end{subequations}
where 
\begin{equation}
a=\frac{R}{\sqrt{q-R^{2}}};\qquad\qquad b=\sqrt{\frac{q}{1-q}};\qquad\qquad v=\sqrt{\frac{q-R^2}{1-q}}.\label{eq:RSabv}
\end{equation}

\subsection{1-RSB free entropy\label{app:1RSB}}
In this case, there are three order parameters $R,q_1,q_0$ (and their conjugates $\hat{R}, \hat{q}_1, \hat{q}_0$) with $q_0\geq q_1$ (resp. $\hat{q}_0 \geq \hat{q}_1$): the two values of the overlap arise from the usual $1$-RSB structure of the overlap matrix $\mathcal{\bm Q}=\left\{q_{ab}\right\}$~\citep{mvpbook}. There is an additional order parameter $\theta$ that tunes the relative weight of the two overlaps, so that the overlap distribution follows $P\left(\left \langle q_{ab}\right \rangle \right)= \theta \delta \left( \left \langle q_{ab}\right \rangle - q_0 \right) + \left(1-\theta\right) \delta \left( \left\langle q_{ab} \right\rangle- q_1\right)$.  The free entropy as a function of these order parameters  ($R,q_1,q_0,\hat{R},\hat{q}_1,\hat{q}_0$) and the corresponding self-consistent equations read:
 \begin{align}
 \mathcal{G} &=-R\hat{R}+\frac{1}{2}\hat{q}_0\left(q_0 -1 \right) -\frac{\theta}{2}\left(q_0\hat{q}_0-q_1\hat{q}_1\right)+\frac{2\alpha}{\theta}\int\mathcal{D}t\,\Phi\left(at\right)\log\int\mathcal{D}\tau\Xi^{\theta}+\frac{1}{\theta}\int\mathcal{D}\tau\log \int \mathcal{D}z 2^{\theta}\text{\ensuremath{\cosh}}^{\theta} \chi \label{eq:RSB_free_energy}
\\
 \Xi &= \Phi\left(bt+c\tau\right)+e^{\frac{\beta^{2}\left(1-p\right)}{2}-\beta\sqrt{1-q_0}\left(bt+c\tau\right)}\Phi\left(\beta\sqrt{1-q_0}-bt-c\tau\right) \label{eq:Xi_1RSB}
\\
  \chi & = \hat{R}+ \sqrt{\hat{q}_0-\hat{q}_1}z + \sqrt{\hat{q}_1}\tau \label{eq:chi_1RSB}
\end{align}
and
\begin{subequations}
\begin{alignat}{2}
R & = \int \mathcal{D}\tau\,\frac{\int \mathcal{D}z\,\text{\ensuremath{\tanh}}\chi\,\text{cosh}^{\theta} \chi}{\int \mathcal{D}z\,\text{cosh}^{\theta} \chi}  ;\quad & \quad \hat{R} & = \sqrt{\frac{2}{\pi}}\alpha\beta\sqrt{\frac{q_0}{q_1-R^{2}}}\int\mathcal{D}te^{-\frac{a^{2}t^{2}}{2}}\frac{\int\mathcal{D}\tau\mathcal{T}\Phi\left(\beta\sqrt{1-q_0}-bt-c\tau\right)\Xi^{\theta-1}}{\int\mathcal{D}\tau\Xi^{\theta}};\\
q_1 & = \int \mathcal{D}\tau\frac{\left(\int \mathcal{D}z\,\text{\ensuremath{\tanh}}\chi\,\text{cosh}^{\theta} \chi\right)^2}{\left(\int \mathcal{D}z\,\text{cosh}^{\theta} \chi\right)^2} \, ; & \hat{q}_1 & =2\alpha\beta^{2}\int\mathcal{D}t\,\Phi\left(at\right)\frac{\left[\int\mathcal{D}\tau\Xi^{\theta-1}\mathcal{T}\Phi\left(\beta\sqrt{1-q_0}-bt-c\tau\right)\right]^{2}}{\left[\int\mathcal{D}\tau\Xi^{\theta}\right]^{2}};\\
q_0 & = \int \mathcal{D}\tau\, \frac{\int \mathcal{D}z\,\text{\ensuremath{\tanh}}^2\chi\,\text{cosh}^{\theta} \chi}{\int \mathcal{D}z\,\text{cosh}^{\theta} \chi} \, ; & \hat{q}_0 & =2\alpha\beta^{2}\int \mathcal{D}t\frac{\Phi\left(at\right)}{\int \mathcal{D}\tau\Xi^{\theta}}\int \mathcal{D}\tau\mathcal{T}^{2}\Xi^{\theta-2}\Phi^{2}\left(\beta\sqrt{1-q_0}-bt-c\tau\right),
\end{alignat}\label{eq:selfconsistentRSB}
\end{subequations}

where  $\mathcal{T} = e^{\frac{\beta^{2}\left(1-q_0\right)}{2}-\beta\sqrt{1-q_0}\left(bt+c\tau\right)}$ and
\begin{equation}
a=\frac{R}{\sqrt{q_1-R^{2}}};\qquad\qquad b=\sqrt{\frac{q_1}{1-q_0}};\qquad\qquad c=\sqrt{\frac{q_0-q_1}{1-q_0}}.\label{eq:RSBabc}
\end{equation}
Note that the structure of both the free entropy and the self-consistent equations (in particular the block related to the energetic term) are almost equivalent to the ones shown in Appendix~\ref{app:MFTheory} for the coupled system with $y$ students (see in particular Eqs. \eqref{eq:free_en_RRS_nodiag} and \eqref{subeq:SP_equations_coupled_Rpq}), provided the mapping $p\leftrightarrow q_0$, $q\leftrightarrow q_1$ (and the same for their conjugates) and $y\leftrightarrow \theta$, although the interpretation of the two parameters $y\in \mathbb{Z}_+$ and $\theta \in \left[0,1\right]$ in the two models is completely different and an additional ferromagnetic interaction $\gamma$ is present in the first case. Given this mapping, also the left equations in \eqref{eq:selfconsistentRSB} are equivalent to \eqref{eq:SPeq_R}-\eqref{eq:SPeq_q}-\eqref{eq:SPeq_p} upon a  Hubbard-Stratonovich transform. \\

\subsubsection{Dynamic Transition line\label{app:Tdyn}}

The dynamic 1-RSB transition temperature $T_{\text{dyn}} \left( \alpha \right)$ can be found as the largest temperature at which the 1-RSB free entropy admits a local maximum at $\theta\to 1$ with $q_0 > q_1$. We recall the procedure discussed in \citep{10.21468/SciPostPhys.2.3.019} that we used also in this case. We start from the 1-RSB free entropy \eqref{eq:RSB_free_energy} and we expand it around $\theta=1$ to the first order:
\begin{equation}
    \mathcal{G}^{\text{1RSB}}\left(R,\hat{R}q_{1},q_{0},\hat{q}_1,\hat{q}_0,\theta\right)=\mathcal{G}^{\text{RS}}\left(R,q_{1},\hat{R},\hat{q}_{1}\right)+\left(\theta-1\right)\left.\partial_{\theta}\mathcal{G}^{\text{1RSB}}\right|_{\theta=1}+O\left[\left(\theta-1\right)^{2}\right].\label{eq:G1RSB_expansion_theta_1}
\end{equation}
 In particular, the 1-RSB free entropy computed at $\theta=1$ gives the RS expression by construction with $q_0=q_1$. On the other hand, the first-order contribution $\delta \mathcal{G}= \left.\partial_{\theta}\mathcal{G}^{\text{1RSB}}\right|_{\theta=1}$ is given by
\begin{align}
\delta\mathcal{G}=\frac{\partial\mathcal{G}^{\text{1RSB}}}{\partial\theta}|_{\theta=1} =& -\frac{1}{2}\left(q_{0}\hat{q}_{0}-q_{1}\hat{q}_{1}\right)-\mathbb{E}_{w^{0}}\int D\tau\log\int Dz\text{cosh}\chi+\mathbb{E}_{w^{0}}\int D\tau\frac{\int Dz\text{cosh}\chi\log\text{cosh}\chi}{\int Dz\text{cosh}\chi}\\
 & -2\alpha\int Dt\Phi\left(at\right)\log\int Dz\Xi+2\alpha\int Dt\Phi\left(at\right)\frac{\int Dz\Xi\log\Xi}{\int Dz\Xi}\label{eq:G1rsb_dertheta_attheta1}
\end{align}
The procedure is to find the global maximum of the $0-$th order term (i.e. the RS part); then, at fixed ($R,q_1$) we compute the equilibrium value of the first-order correction \eqref{eq:G1rsb_dertheta_attheta1} through another saddle-point w.r.t. $q_0$. The latter order parameter and its conjugate $\hat{q}_0$ are fixed by the following saddle point equations:
\begin{align}
    q_0 &= \int D\tau\frac{\int Dz\text{tanh}\chi\text{sinh}\chi}{\int Dz\text{cosh}\chi}\\
    \hat{q}_0 &= 4\alpha\int Dt\Phi\left(at\right)\left[\frac{\int Dz\left(\partial_{q_{0}}\Xi\right)\log\Xi}{\int Dz\Xi}-\left(\frac{\int Dz\Xi\log\Xi}{\int Dz\Xi}\right)\left(\frac{\int Dz\partial_{q_{0}}\Xi}{\int Dz\Xi}\right)\right]
\end{align}
where $\chi, \Xi$ are given by Eqs.~\eqref{eq:chi_1RSB}-\eqref{eq:Xi_1RSB}, respectively.

\section{Simulated Annealing's implementation details and additional numerical results \label{app:SAdetails}}
In this section we discuss the implementation parameters of the simulated
annealing (SA). The model \eqref{eq:HamiltonianCoupledPerceptrons} is initialized at a temperature $T_{0}=0.4$,  and each student's weight vector $\boldsymbol{w}_u$ is drawn at random from $\{-1,1\}^N$, independently on the
others. At each temperature we perform $1$ MonteCarlo sampling sweep, where a move is proposed for every degree of freedom $w_{i,u}$, and it is accepted/rejected according to the Metropolis choice. Then the temperature is
linearly decreased so that $T_{l}=T_{l-1}-\eta$ with $\eta$ a suitable annealing rate, and the sampling is repeated starting from the last configuration at the previous (higher) temperature. The annealing process continues up to a final temperature $T_f=10^{-2}$. In all the simulations shown performed in this work, we used $\eta \in 10^{\{-2,-3,-4,-5,-6\}}$ (for computational time reasons, the latter value has been used only for $y=1$, see Fig.~\ref{fig:RSA_app_effect-eta}). 
The energy shifts can be efficiently computed using the same procedure discussed in \citep{baldassi_unreasonable_2016} (Supporting Information). 

\section{Approximate Message Passing \label{app:AMP}}

\begin{figure}[t!]
    \centering
    \includegraphics[width=0.7\textwidth]{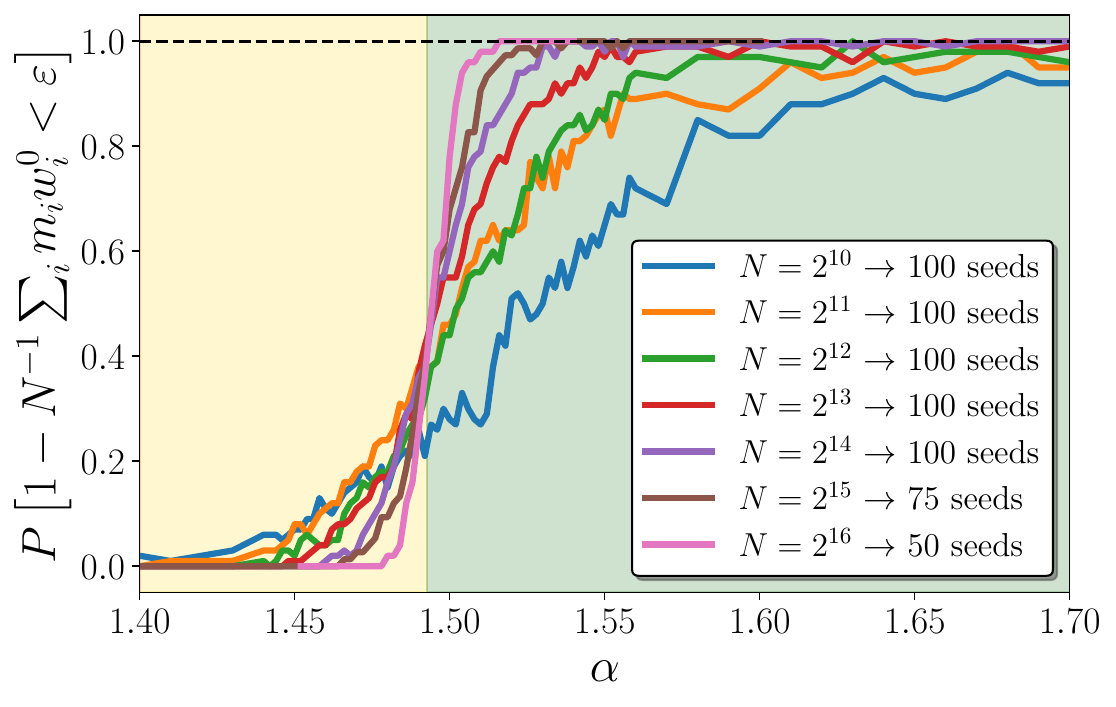}
    \caption{AMP's empirical probability of finding the teacher configuration vs the fraction of examples $\alpha$. Each curve corresponds to a different system size $N$ (shown in the caption) and shows the amount of instances for which AMP converges to the teacher, i.e. to a set of marginals $\boldsymbol{m}$ such that $1-\sum_i{m_i w_i^0} \slash N <\varepsilon$ where $\boldsymbol{w^0}$ is the teacher configuration and $\varepsilon$ is the convergence threshold on the AMP marginal probabilities used for these experiments (here $\varepsilon=10^{-7}$). The simulations are done at the Bayes optimal temperature $T=0$.}
    \label{fig:AMP}
\end{figure}

In this section we discuss some numerical experiments on the Approximate Message Passing (AMP) algorithm that we used to check the spinodal point at $T=0$, i.e. the hard-easy boundary as found by the RS theory. We do not discuss the details of the method itself, as it has been extensively analyzed in several previous works: in particular, the AMP implementation we used here is taken from the general derivation for Generalized linear estimation problems (GLMs, the binary perceptron in the teacher-student scenario being an example of) presented in the review article \citep{zdeborova2016statistical} (Section VI-C). Similarly to the SA experiments, we run the AMP algorithm by fixing the system size $N$ and varying the amount of examples $\alpha=M\slash N$. For each instance, AMP is run until convergence is reached on the marginal probabilities for each weight (with a finite tolerance $\varepsilon=10^{-7}$).  We then compute the empirical probability (averaging over all the instances) that the final overlap between the teacher configuration and the AMP's marginal is $1$. Results are shown in Fig. \ref{fig:AMP} for different system sizes $N$. We can clearly see how the curves get sharper and sharper by increasing $N$ and the transition to a perfect recovery approaches the boundary between the hard and the easy phases as computed from the RS phase diagram at $T=0$, so that $\alpha_{\text{AMP}} = \alpha_c$.

 \begin{figure}[h!]
\includegraphics[width=\textwidth]
{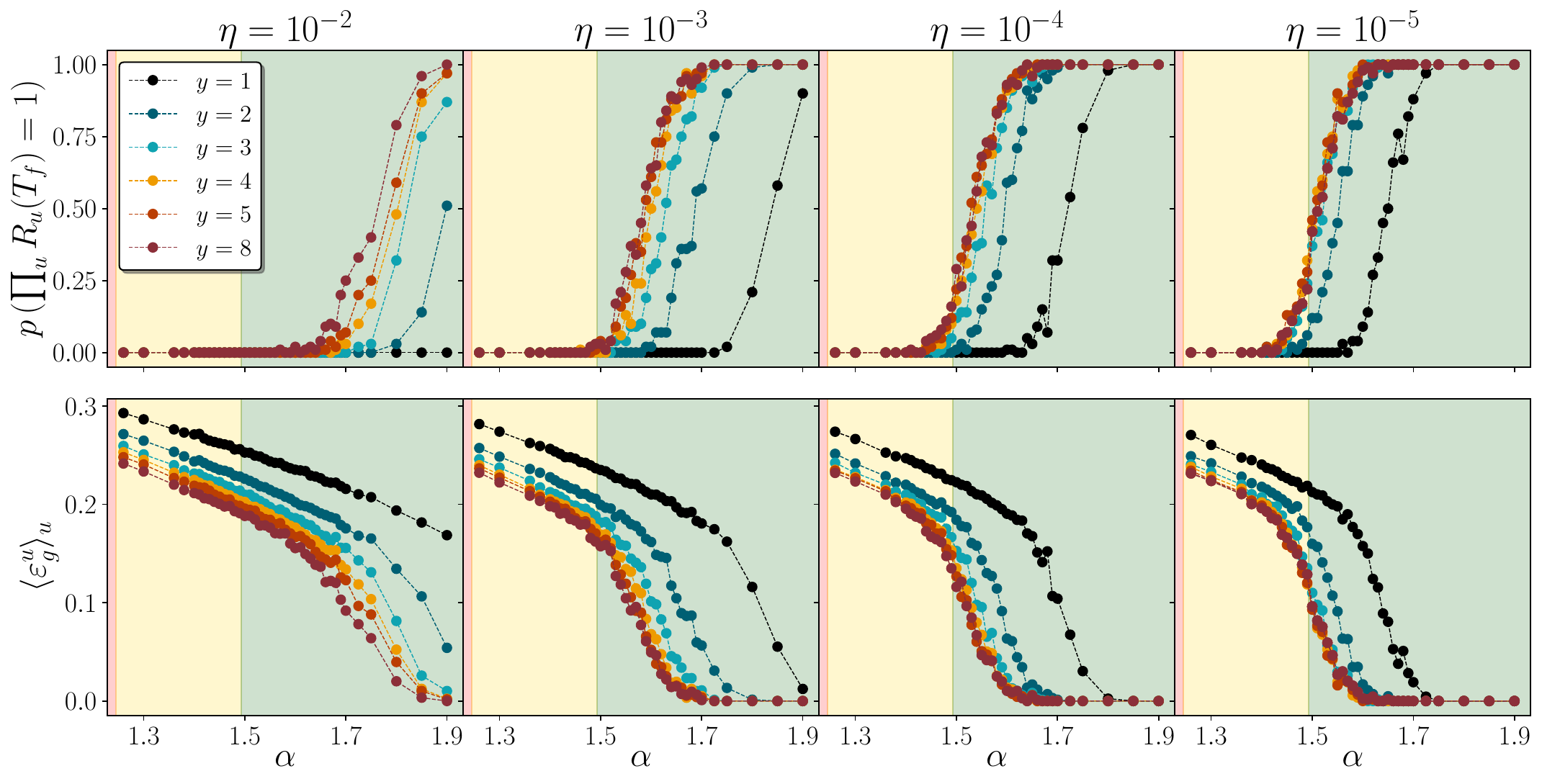}
    \caption{Numerical performances of SA/RSA: the top row shows the empirical probability that the teacher configuration is found at the end of the SA, the bottom row shows the mean generalization error; both quantities are plotted vs of the fraction of samples $\alpha$, for a system with $N=2001$ weights, for different values of the annealing rate $\eta$ (shown at the top of each column). Comparison between the single non-replicated Perceptron (black lines) and a system of $y$ coupled Perceptrons with different values of $y$, with fixed coupling $\gamma=1$. For $y\geq 2$, the top row shows the probability that \textit{all} the $y$ students find the teacher configuration at the end of the SA, and the bottom row displays the mean generalization error averaged also on the $y$ students. However, we numerically observe that, except at extremely fast annealing rates (e.g the first panel with $\eta=10^{-2}$) the students display a practically identical behavior in temperature.  The right-most panels are the same shown in Fig.~\ref{fig:res_RSA} of the main text.
    \label{fig:RSA_app}}
\end{figure}
\begin{figure}[h!]
    \includegraphics[width=\textwidth]
    {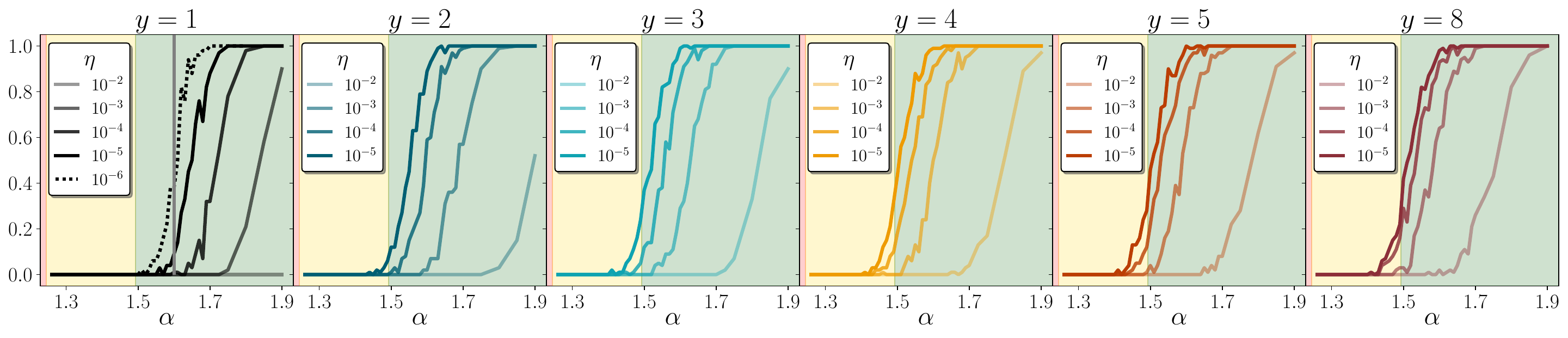}
  \caption{Same results as in Fig.~\ref{fig:RSA_app} but this time each panel refers to one value of $y$: each panel shows the probability of finding the teacher solution at the end of the SA for different values of $\eta$: we can observe how the cooling rate seems to converge to the same curves in the case $y=5$ and $8$. The left-most panel ($y=1$) shows also the result obtained with an annealing rate $\eta=10^{-6}$ over $50$ seeds (dotted line). The vertical gray line in the left-most panel shows the critical value of $\alpha$ at which $1$-RSB glassy states disappear, $\alpha_c^{\text{RSB}}\approx 1.6$. The settings are identical to those of Fig.~\ref{fig:RSA_app}. \label{fig:RSA_app_effect-eta}}
\end{figure}

\end{document}